\documentclass[twocolumn]{jpsj2}
\def\v#1{\mib #1}

\def\Ms{M_{\rm s}}

\newcommand{\ket}[1]{\left\vert {#1} \right\rangle}

\newcommand{\aver}[1]{\left\langle {#1} \right\rangle}
\def\runauthor{ Kazuo {\sc Hida}}
\title
{
Ferrimagnetic  and  Long Period Antiferromagnetic Phases in  High Spin  Heisenberg Chains with $D$-Modulation}

\author
{
Kazuo {\sc Hida}\thanks{E-mail: hida@phy.saitama-u.ac.jp}
}

\inst
{Division of Material Science, Graduate School of Science and Engineering, Saitama University, Saitama, Saitama, 338-8570
 
}

\recdate
{
\today
}

\abst
{
The ground state properties of the high spin Heisenberg chains with alternating single site anisotropy are investigated by means of the numerical exact daigonaization and DMRG method. It is found that the ferrimagnetic state appears between the Haldane phase and period doubled N\'eel phase for the integer spin chains. On the other hand, the transition from the Tomonaga-Luttinger liquid state  into the ferrimagnetic state takes place for the half-odd-integer spin chains. In the ferrimagnetic phase, the spontaneous magnetization varies continuously with the modulation amplitude of the single site anisotropy. Eventually, the magnetization is locked to fractional values of the saturated magnetization. These fractional values satisfy the Oshikawa-Yamanaka-Affleck condition. The local spin profile is calculated to reveal the physical nature of each state. In contrast to the case of frustration induced ferrimagnetism, no incommensurate magnetic superstructure is found.
}

\kword
{
high spin Heisenberg chain, single-site anisotropy, spatial modulation,  Haldane phase, ferrimagnetic phase, period-doubled N\'eel phase, hidden antiferromagnetic order
}

\begin{document}
\sloppy
\maketitle
\section{Introduction}

Among various exotic ground states in quantum magnetism, the Haldane state in the integer spin antiferromagnetic Heisenberg chain\cite{fd} has been most extensively studied both experimentally and theoretically. This state is characterized by the hidden antiferromagnetic string order accompanied by the $Z_2\times Z_2$ symmetry breakdown in spite of the presence of the energy gap and exponential decay of the spin-spin correlation function. The easy plane single-site anisotropy $D (>0)$ destroys the Haldane ground state leading to the large-$D$ state with finite energy gap and  exponential decay of the spin-spin correlation function {\it without specific order}. On the contrary, the easy axis  single-site anisotropy ($D <0$) drives the Haldane state into the N\'eel state.\cite{md,ht,chen} 

On the other hand, the ground state of the odd spin Heisenberg chain is the Tomonaga-Luttinger liquid state. Due to its critical nature, the ground state is driven to the N\'eel ordered state for infinitesimal negative $D$ while the Luttinger liquid state is stable against positive $D$\cite{hjs}.

In this context, it is an interesting issue to investigate how the ground states of the quantum spin chains are modified if the  easy-axis and easy-plane single-site anisotropy coexist in a single chain. In the previous work\cite{altd}, the present author and Chen  investigated the $S=1$ chain with alternating single-site anisotropy and found that the period doubled N\'eel phase with $\ket{\uparrow 0 \downarrow 0}$ structure is realized for strong alternation amplitude, although the Haldane phase is stable for weak alternation. The physical origin of this type of N\'eel order is interpreted as a 'pinning' of the string order. In the present work, we further explore this problem for the cases $S > 1$. We find not only the  $\ket{\uparrow 0 \downarrow 0}$ ground state but also the ferrimagnetic ground states with quantized and unquantized spontaneous magnetization for intermediate strength of $D$-alternation. These quantizated values of magnetization also satisfy the Oshikawa-Yamanaka-Affleck condition\cite{oya} well-known for the magnetization plateau in the magnetic field.

This paper is organized as follows. In the next section, the model Hamiltonian is presented and the two possible senarios which leads to  different ground states are explained. In \S 3, the numerical results for the spontaneous magnetization and the  local spin profile are presented to reveal the physical nature of each state. The last section is devoted to summary and discussion.

\section{Model Hamiltonian}

We investigate the ground state of the  Heisenberg chains with alternating single site anisotropy whose Hamiltonian is given by,
\begin{eqnarray}
\label{ham0}
{\cal H} &=& \sum_{l=1}^{N}J\v{S}_{l}\v{S}_{l+1}+\delta D\sum_{l=1}^{N/2}S_{2l-1}^{z2}\nonumber\\
&-&\delta D\sum_{l=1}^{N/2}S_{2l}^{z2}, \ \ (J > 0, \delta D >0).
\end{eqnarray}
where  $\v {S_{i}}$ is the spin-$S$ operator on the $i$-th site.  

In ref. \citen{altd}, it is found that the period doubled N\'eel phase with $\ket{\uparrow 0 \downarrow 0}$ structure is realized for large enough $\delta D$ for $S=1$, although the Haldane phase is stable for small $\delta D$. The mechanism to stabilize this period doubled N\'eel phase can be understood along the following senario (senario I). In the absence of the $D$-terms, the $S=1$ ground state has a hidden string order which implies that the spins with $\ket{\pm 1}$ are arranged antiferromagnetically if the sites with $\ket{0}$ are skipped.\cite{md,ht} The position of the sites with $\ket{\pm 1}$ and  $\ket{0}$ strongly fluctuate quantum mechanically and this antiferromagnetic order remains hidden because it is impossible to observe the correlation  between only the sites with  $\ket{\pm 1}$ experimentally. In the presence of strong $\delta D$-terms, only the states consistent with the constraint set by these $\delta D$-terms survive  among all states with hidden order.  For $\delta D>>J$, the odd-th site must be in the state $\ket{0}$ and the even-th sites $\ket{\pm 1}$. To be compatible with the string order, spins must be arranged as $\ket{\uparrow 0 \downarrow 0}$. Thus the strong $\delta D$-term can select the spin states among those with hidden order to realize the explicit period doubled N\'eel order.

\begin{figure}
\centerline{\includegraphics[width=30mm]{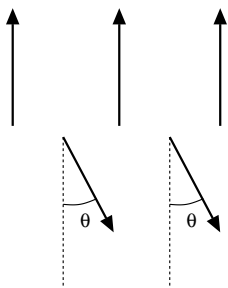}}
\caption{The classical spin configuration in the ferrimagnetic state}
\label{claspin}
\end{figure}

\begin{figure}
\centerline{\includegraphics[width=70mm]{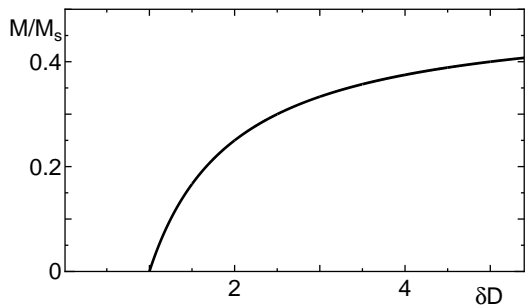}}
\caption{The spontaneous magnetization in the classical limit. Here and in the following figures energy scale is set as $J=1$.}
\label{clasmag}
\end{figure}

On the other hand,  another  senario (senario II) is possible from classical point of view, although this is not realized in the $S=1$ case. Let us consider the classical limit where each spin can be regarded as a classical unit vector. In the ground state, the spins on the easy axis sites are fully polarized along $z$-direction but those on the easy-plane sites are tilted by an angle $\theta=\cos^{-1}(J/\delta D)$ from the $-z$-direction as shown in Fig. \ref{claspin}. Therefore this senario leads to the noncollinear ferrimagnetic ground state with the spontanenous magnetization $M=M_{\rm s}(1-(J/\delta D))/2$ as plotted in Fig.\ref{clasmag}. 

In what follows, we show that either of these two senarios can be realized in $S >1$ chains depending on the values of $\delta D$ and $S$ based on the numerical exact diagonaliztion (NED) and density matrix renormalization group (DMRG) calculation.

\section{Numerical Results}
\subsection{Spontaneous Magnetization}

To identify the ferrimagnetic regime expected in senario II, the ground state spontaneous magnetization is calculated by NED with periodic boundary condition and by DMRG with open boundary condition for various values of $S$ and $\delta D$. The maximum chain length for the NED is $N=12$ for $S=3/2$, $S=2$ and $S=5/2$, while it is  $N=8$ for $S=3$. For the DMRG calculation with open boundary condition, appropriate end spins are added to reduce the boundary effects. 

\begin{figure}
\centerline{\includegraphics[width=70mm]{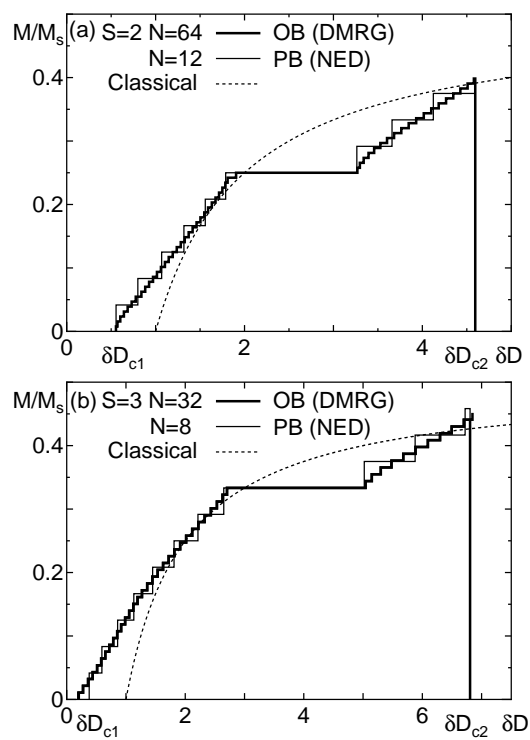}}
\caption{The spontaneous magnetization for (a) $S=2$ with $N=12$ (NED) and $N=64$(DMRG), and (b) $S=3$ with $N=8$ (NED) and $N=32$ (DMRG) plotted against $\delta D$. The dotted lines are the classical spontaneous magnetization.}
\label{mageven}
\end{figure}

\begin{figure}
\centerline{\includegraphics[width=70mm]{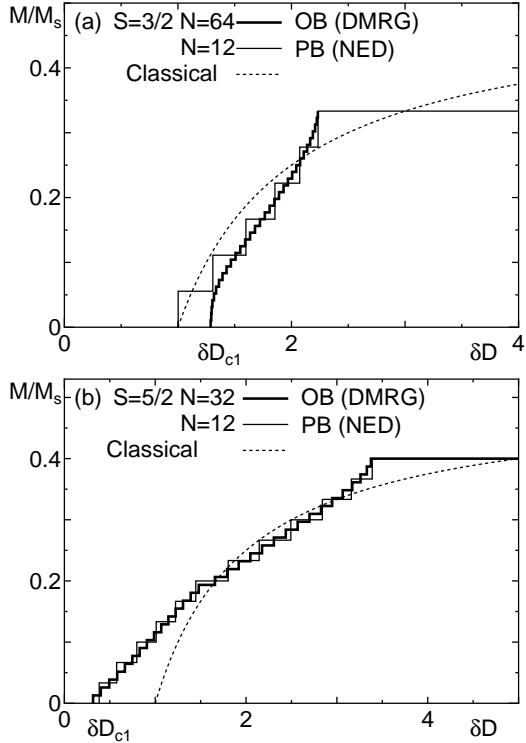}}
\caption{The spontaneous magnetization for (a) $S=3/2$ with $N=12$ (NED) and $N=64$(DMRG) and  (b)  $S=5/2$ with $N=12$ (NED) and $N=32$(DMRG) plotted against $\delta D$. The dotted lines are the classical spontaneous magnetization.}
\label{magodd}
\end{figure}

The results for the integer spin cases are presented in Fig. \ref{mageven} for $S=2$ and 3 and those for the half-odd-integer spin cases are presented  in  Fig. \ref{magodd} for $S=3/2$ and $5/2$. In contrast to the case of $S=1$, it is found that the ferrimagnetic phase always appear for $S \ge 3/2$ above the critical value $\delta D_{\rm c1}$. For $0 < \delta D < \delta D_{\rm c1}$,  the energy gap decreases monotonously with $\delta D$ until it vanishes at $\delta D=\delta D_{\rm c1}$ in all cases studied ($3/2 \leq S \leq 3$). Therefore, we may safely conclude that the ground state is the Haldane phase or the Tomonaga-Luttinger liquid phase according as $S=$ integer or half-odd-unteger. 

For integer $S$, the spontaneous magnetization vanishes for large enough $\delta D$. From the local spin profile $\aver{S^z_i}$ which will be presented in the next section, this state turns out to be the period-doubled $\ket{\uparrow 0 \downarrow 0}$-type N\'eel state expected in senario I. On the other hand, the ferrimagnetic state remains stable for arbitrarily large $\delta D$ for half-odd-integer $S$ because the ground state of the easy plane site is  the doublet $S^z=\pm 1/2$ which can sustains the  ferrimagnetic order even for large $\delta D$. 

It should be also noted that the spontaneous magnetization in the ferrimagnetic phase is not restricted to simple fractions of the saturation magnetization $M_{\rm s}(=NS)$ as in the usual quantum ferrimagnets\cite{yama1,yama2} but varies  continuously with $\delta D$ in accordance with the classical intuition predicting the noncollinear ferrimagnetism in senario II. 

Within a appropriate range of $\delta D$, however, the spontaneous magnetization is locked to a simple fraction of $M_{\rm s}$ reflecting the quantum nature of the present model. In the DMRG calculation, these quantized value of magnetization  slightly deviates from the simple fraction of $M_{\rm s}$ due to the boundary spins. Correspondingly,  the spontaneous magnetization is slightly rescaled so that the main quantized value exactly equals the simple fraction of $M_{\rm s}$ in Figs. \ref{mageven} and \ref{magodd}.

\begin{figure}
\centerline{\includegraphics[width=70mm]{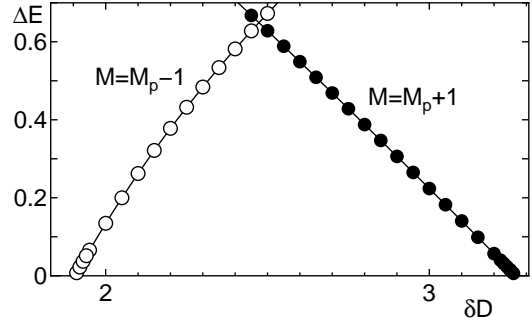}}
\caption{The energy gap of $S=2$ chain on the plateau state with magnetization $M_{\rm p}=M_{\rm s}/4$ for $N=72$. The filled (open) symbols are the gap to the state with magnetization $M=M_{\rm p}+1$($M_{\rm p}-1$) }
\label{pla2}
\end{figure}

In all cases, these quantized values of the magnetization satisfy the condition
\begin{equation}
p(S-m)=q
\label{oymcond}
\end{equation}
where $p$ is the size of the unit cell, $q$ is an integer and $m$ is the magnetization per site ($m=M/N=MS/\Ms$).  In the present model, $p$ is equal to 2. This condition is identical to that proposed by Oshikawa, Yamanaka and Affleck\cite{oya} for the magnetization plateau in the magnetic field. However, their proof is not restricted to the magnetic field induced magnetization but also applies for the spontaneous magnetization in the ferrimagnetic phase. If the condition (\ref{oymcond}) is satisfied, it is allowed to have a finite energy gap to the excited state with different magnetization. This implies the stability of the ground state against the variation of $\delta D$ which leads to  the 'plateau' behavior. The $\delta D$-dependence of the energy gap on the plateau state calculated by the DMRG method for $S=2$ chains is shown in Fig. \ref{pla2}. It is clear that the energy gap is finite on the plateau region $1.91 \leq \delta D \leq 3.26$.

In addition, the maximum possible values of the ground state spontaneous magnetization is bounded from above due to the nature of the present model. To maximize the spontaneous magnetization, the spins on the easy axis sites must have $S^z_i=S$ and those on the easy plane site must have negative value  due to the antiferromagnetic interaction with the neighbouring polarized spins.  It should be noted that the senario I leading to the period doubled N\'eel order becomes effective if $S^z_i=0$ on the easy plane site.  Therefore the spontaneous magnetization per site $m$ must satisfy $0 < m = MS/\Ms < S/2$ in the ferrimagnetic phase. This implies $2S > q > S$.  In the half-odd-integer case, the smallest possible value of $q$ is  $S+1/2$. This gives the magnetization per site $m=(S-1/2)/2$ which yields $M/\Ms=(S-1/2)/2S$. On the other hand, in the integer spin case, the smallest possible value of $q$ is $S+1$. In this case, $m$ is equal to $(S-1)/2$ which yields $M/\Ms=(S-1)/2S$. It should be noted this value vanishes for $S=1$. This explains why the quantized ferrimagnetic phase does not appear for $S=1$ case.

\begin{figure}
\centerline{\includegraphics[width=70mm]{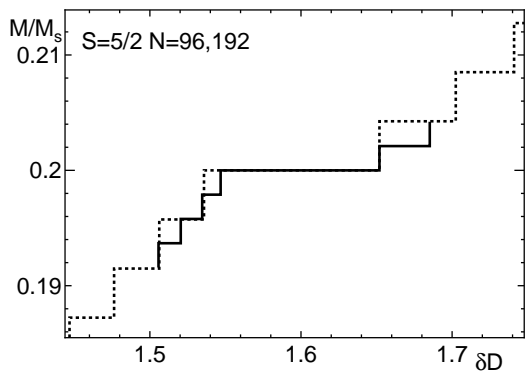}}
\caption{The small 'plateau' at spontaneous magnetization $M=M_{\rm s}/5$ for  $S=5/2$  plotted against $\delta D$. The system size is $N=96$ (dotted line) and 192 (solid line). Rescaling factor is slightly different from that of Fig. \ref{magodd} to fix this small plateau precisely to $M=M_{\rm s}/5$. }
\label{mag5ov2lar}
\end{figure}

Actually, prominent 'plateaus' are observed only for the smallest possible value of $q$.  This is due to the fact that the condition  for the gap generation on the compactification radius of the underlying Gaussian model becomes increasingly severer with the increase of $q$\cite{oya}. As a secondary plateau with larger $q$,  we only find a small plateau at $M=\Ms/5$ for $S=5/2$ which corresponds to  $q=4(=S+3/2)$ as shown in Fig. \ref{mag5ov2lar} within the $S$ values studied so far, although this plateau is almost unvisible for small sized systems shown in Fig. \ref{magodd}(b).
\subsection{Local Magnetization Profile}

\begin{figure}
\centerline{\includegraphics[width=70mm]{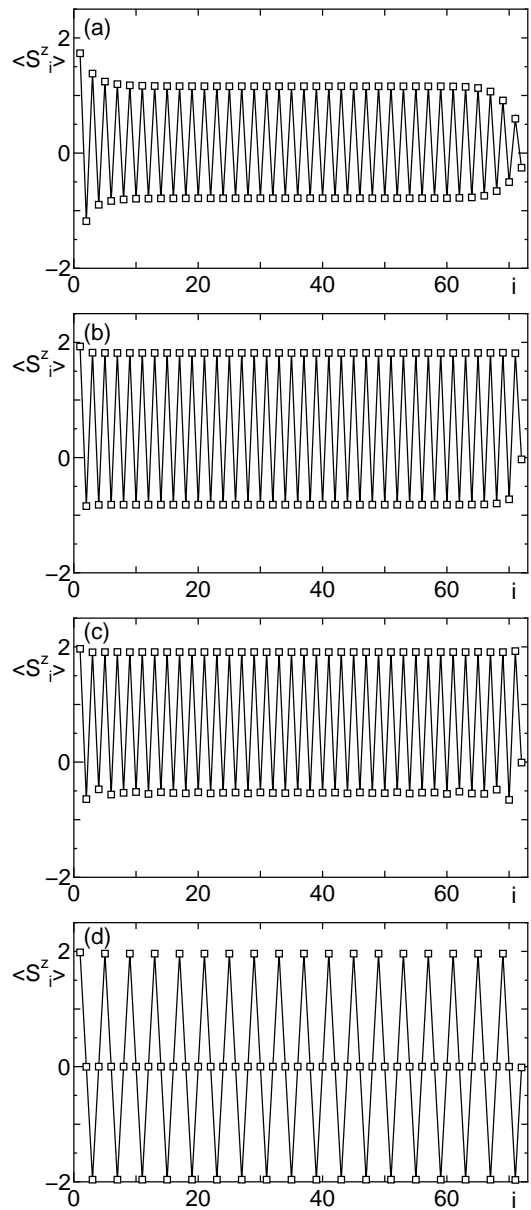}}
\caption{The local magnetization profile of $S=2$ chains for (a) $\delta D=1.0$ (below the plateau),  (b) $2.5$ (on the plateau), (c) 4.0 (above the plateau) and (d) $5.0$ (period-doubled N\'eel phase) with $N=72$.}
\label{cor}
\end{figure}

The local magnetization profile $\aver{S^z_i}$ calculated by the DMRG method is presented for each phase of $S=2$ chains  in Fig. \ref{cor}. Below the plateau, the easy plane spins are almost in the state $S^z_i=-1$ while the magnetization of the easy axis spins increases from 1 to 2 as $\delta D$ approaches the lower end of the plateau region. On the plateau, the easy axis spins are almost in the state $S^z_i=2$ and the easy plane spins are in the state $S^z_i=-1$ leading to the quantized value of spontaneous magnetization for the smallest possible value of $q$ described in the preceding section. Above the plateau, the easy axis spins are almost in the state $S^z_i=2$ and the increase in total spontaneous magnetization is due to the decrease in the polarization of the easy plane spins. 

The behavior of the local magnetization profile in the noncollinear ferrimagnetic phase is in contrast to the similar noncollinear ferrimagnetic phase in the frustrated spin chains investigated in refs. \citen{ym1} and \citen{zigferri} in which the  incommensurate superstructure is observed.  This suggests that the incommensurate superstructures observed in these literatures are essentially due to frustration.

In these ferrimagnetic phases, the correlation between the easy axis spins are ferromagnetic. At $\delta D=\delta D_{\rm c2}$, however, the easy plane spins turns into the state with $S^z_i=0$ and the correlation between the easy axis spins turns into antiferromagnetic. In this case, the magnetion profile clearly shows the $\ket{\uparrow 0 \downarrow 0}$ structure as shown in Fig. \ref{cor}(d). For the calculation of the local spin profile in this phase, we have applied a tiny symmetry breaking field  with period 4, because otherwise the true ground state of the finite size system is the linear combination of $\ket{\uparrow 0 \downarrow 0}$ and $\ket{\downarrow 0 \uparrow 0}$-type states and no net local magnetization is expected. Actually, the DMRG calculation is often trapped to the states with domain walls in the absence of the symmetry breaking field. The value of the symmetry breaking field ranged from $0.005$ to $0.02$ and the results turned out to be almost indistinguishable on the scale of the Fig. \ref{cor}(d).  The physical origin of this magnetic structure is understood in the same way as the $S=1$ case\cite{altd} following the first senario described in \S 2.

\section{Summary and Discussion}

The ground state properties of the high spin Heisenberg chains with alternating single site anisotropy are investigated by means of the numerical exact daigonaization and DMRG method. It is found that the ferrimagnetic state appears between the Haldane phase and period doubled N\'eel phase for the integer spin chains. On the other hand, the transition from the Tomonaga-Luttinger liquid state  into the ferrimagnetic state takes place for the half-odd-integer spin chains. In the ferrimagnetic phase, the spontaneous magnetization varies continuously with the modulation amplitude of the single site anisotropy in accordance with the classical intuition. Eventually, however, the magnetization is locked to fractional values of the saturated magnetization which satisfies the Oshikawa-Yamanaka-Affleck condition. 

The local spin profile is calculated to reveal the physical nature of each state. In contrast to the case of frustration induced ferrimagnetism\cite{ym1,zigferri}, no incommensurate superstructure is found. We thus expect that the incommensurate superstructures found is these literatures are essentially due to the interplay of quantum effect and frustration.

The similar mechanism should also work in 2 and 3 dimensions, although the Haldane or Tomonaga Luttinger liquid phase would be replaced by the long range ordered N\'eel-type state. However, even in the large $\delta D$ limit, the ground state is not trivial due to the frustration in the effective interaction among the easy-axis spins. The investigation of these higher dimensional models are in progress. 

For the experimental realization of the present mechanism it is necessary to synthesize the compound of easy axis magnetic ions and easy plane ones. Considering a variety of phases expected for the present model, this is a challenging attempt for the experimentalists. Recently, various single chain molecular magnets with considerable strength of single site anisotropy have been synthesized.\cite{scm1,scm2} Although the materials with alternating sign $D$-terms with uniform $S$ are not yet reported, these series of materials can be a good candidate to observe the phenomena proposed in the present work.

The computation in this work has been done using the facilities of the Supercomputer Center, Institute for Solid State Physics, University of Tokyo and  the Information Processing Center of Saitama University.  The diagonalization program is based on the TITPACK ver.2 coded by H. Nishimori. This works is supported by the Grant-in-Aid for Scientific Research from the Ministry of Education, Culture, Sports, Science and Technology, Japan.

\end{document}